\documentclass[aps,prb,a4paper,amsmath,twocolumn,showpacs,floatfix]{revtex4}
\usepackage{graphicx}
\def\lsim{\mathrel {\vcenter {\baselineskip 0pt \kern 0pt
    \hbox{$<$} \kern 0pt \hbox{$\sim$} }}}

\begin{document}

\title{Effect of particle geometry on phase transitions in two-dimensional liquid crystals}

\author{Yuri Mart\'{\i}nez-Rat\'on}
\email{yuri@math.uc3m.es}

\affiliation{Grupo Interdisciplinar de Sistemas Complejos (GISC), 
Departamento de Matem\'aticas, Escuela Polit\'ecnica Superior, 
Universidad Carlos III de Madrid,
Avenida de la Universidad 30, E-28911 Legan\'es, Madrid, Spain.
}

\author{Enrique Velasco}
\email{enrique.velasco@uam.es}

\affiliation{Departamento de F\'{\i}sica Te\'orica de la Materia Condensada
and Instituto de Ciencia de Materiales Nicol\'as Cabrera,
Universidad Aut\'onoma de Madrid, E-28049 Madrid, Spain.}

\author{Luis Mederos}
\email{l.mederos@icmm.csic.es}

\affiliation{Instituto de Ciencia de Materiales, Consejo Superior de
Investigaciones Cient\'{\i}ficas, E-28049 Cantoblanco, Madrid, Spain.}

\date{\today}

\begin{abstract}
Using a version of density-functional theory which combines Onsager 
approximation and 
fundamental-measure theory for spatially nonuniform phases, we have studied 
the phase diagram of freely rotating hard rectangles and hard discorectangles.
We find profound differences in the phase
behavior of these models, which can be attributed to their different packing
properties. Interestingly, bimodal orientational distribution functions are
found in the nematic phase of hard rectangles, which cause a certain degree of 
biaxial order, albeit metastable with respect to spatially ordered phases. This
feature is absent in discorectangles, which always show unimodal behavior.
This result may be relevant in the light of recent experimental results which
have confirmed the existence of biaxial phases. We expect that some perturbation
of the particle shapes (either a certain degree of polydispersity or even
bimodal dispersity in the aspect ratios) may actually destabilize spatially 
ordered phases thereby stabilizing the biaxial phase.
\end{abstract}

% 64.70.Md  Transitions in liquid crystals
% 64.75.+g  Solubility, segregation, and mixing; phase separation
% 61.20.Gy  Theory and models of liquid structure

\maketitle

\section{Introduction} 
Biaxial liquid-crystalline 
phases were hypothesized in 1970 by Freiser \cite{Freiser}. Since then many theoretical models
have described the conditions that promote biaxial order. However, 
the experimental search for biaxial phases has been unfruitful until very recently, when Madsen et al. \cite{Madsen} and Acharya et al. \cite{Acharya}
presented convincing evidence 
for the existence of biaxial order in a fluid composed of 
boomerang-shaped molecules.
In previous work Schlacken et al. \cite{Schlacken} analyzed the effect of 
particle shape on the orientational
properties of two-dimensional fluids using scaled-particle theory (SPT). They considered hard
ellipses and hard rectangles (HR) and, for the latter, located a spinodal which corresponds to a
tetratic phase, where the fluid has two nematic directors oriented perpendicular to each other.
Simulation works have been performed by Cuesta and Frenkel \cite{Cuesta0} on hard ellipses and by 
Bates and Frenkel \cite{Bates} who focused on hard discorectangles (HDR)
and the character of the orientational transition from the isotropic phase. 
Also Lagomarsino et al. 
\cite{Lagomarsino} found similar behavior in a 
fluid of hard spherocylinders confined between two parallel
walls which, in the limit
of small wall separation, can be considered as a realization of a two-dimensional nematic.
These simulations have 
shown that the character of orientational transitions depends strongly on the particle shape: whereas for hard discorectangles the isotropic-nematic transition is of the Kosterlitz-Thouless (KT) type\cite{Frenkel}
in the whole aspect ratio regime where the nematic phase is stable against the solid phase\cite{Bates}, for 
hard ellipses a tricritical point separates regions 
of continuous from first order orientational 
transition \cite{Cuesta0}. 

The density functional (DF) 
formalism has been proven to be a powerful tool in the 
study of bulk nonuniform phases of 
liquid crystals in three dimensions, specifically for the model of 
freely rotating hard spherocylinders  
\cite{Somoza}$^-$\cite{Velasco}. To our knowledge,
in two dimensions the DF formalism has been applied only to the study of 
uniform liquid-crystalline phases. 
In the present work we will not focus on questions related to the KT nature 
of the isotropic-nematic phase transition,
but rather we investigate by DF theory the occurrence of biaxial
nematic order in fluids and how this feature depends on the geometry of the particles; we consider
both HR and HDR as interaction models.
Also we apply DF theory
to the study of spinodal instabilities to nonuniform phases in order to
analyze the relative stability of these biaxial phases against spatially
ordered phases.

The results from the theory allow us to conclude that a 
certain degree of biaxiality is present in the case of HR. This is
truly remarkable, since this order arises in a one-component system of 
particles with uniaxial symmetry.
However the bifurcation analysis using the DF theory
allow us to conclude that the tetratic phase is metastable 
w.r.t. the solid phase. The remainder of the paper is organized as follows:
In Section II we introduce the theoretical models for both the uniform phases
(SPT in Section II.A) and for the non-uniform ones (DF in Section II.B). 
Subsection II.C is devoted to the bifurcation analysis.
Results are shown in Section III while some conclusions are drawn in Section IV.

\section{Model} 

The DF approximation 
proposed is based on SPT, which has proved to be
a generic theory which is easily extended to any convex particle geometry 
\cite{Cotter,Barboy}. 
Two extensions, suitable for nonuniform phases, are proposed.
Both tend to the SPT in the limit of uniform
phases and take into account ideas from Fundamental Measure theory (FMT).
This theory has been worked out for a system of hard parallelepipeds in 
the restricted orientations approximation 
(the so-called Zwanzig model\cite{Cuesta}) and the corresponding 
phase diagram, including all nonuniform
phases, was recently calculated \cite{Martinez-Raton}. One of the
proposed extensions 
recovers the Onsager second-order virial theory in the 
low-density limit. The other version is built so as to recover the FMT for parallel 
hard rectangles, 
i.e. restricting the particle orientations to point along the nematic 
director. The construction of a functional which interpolates between 
the Onsager limit and the high density one using FMT
for hard spheres has been 
proposed by Cinacchi and Schmid for a system of hard spherocylinders 
in three dimensions\cite{Cinacchi}. Nevertheless 
their proposal seems to be numerically intractable for nonuniform phases. Here we are interested 
in constructing a workable density functional.

\subsection{Uniform phases: Scaled particle theory} 

\begin{figure}[h]
{\centering \resizebox*{5.cm}{!}{\includegraphics{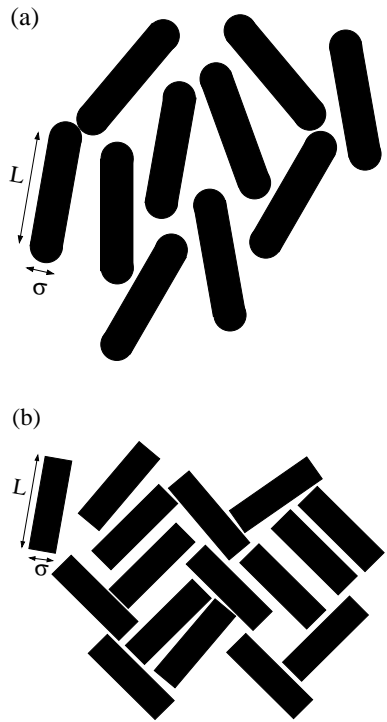}} \par}
\caption{\label{figm1}\small A system of (a) hard discorectangles in
a nematic phase, and (b) hard rectangles in a tetratic phase.
}
\end{figure}
The radically different orientational ordering properties shown by the
two particle models, HDR and HR (see Fig. \ref{figm1}), 
can be understood in the framework of SPT.
A brief discussion of this theory, as applied to our model systems, follows. 

The key quantity to derive the theory is the excluded volume between fluid particles
of length $L$ (HR) or $L+\sigma$ (HDR) and width $\sigma$, 
and a test scaled particle of dimension
$L_s$ and $\sigma_s$ with fixed orientation $\hat{\bf u}_1(\phi_1)$, where $\phi_1$ is the
angle between the uniaxial axis and the nematic director. The excluded volume
is to be averaged over all possible orientations of the fluid particles, 
\begin{eqnarray}
\left<V_{\rm{excl}}^{\alpha}\right>(L_s,\sigma_s,\phi_1)=\int d\phi_2 h(\phi_2) 
V_{\rm{excl}}^{\alpha}(L_s,\sigma_s,\phi_{12}), 
\end{eqnarray}
where $h(\phi)$ is the one-particle orientational distribution function, 
$\phi_{12}=\phi_1-\phi_2$, and 
$V_{\rm{excl}}^{\alpha}$ is the excluded volume between two particles which, for rectangles, 
has the form
\begin{eqnarray}
V_{\rm{excl}}^{\rm{HR}}(L_s,\sigma_s,\phi_{12})&=&
(LL_s+\sigma\sigma_s)|\sin \phi_{12}|+{\cal L}_L{\cal L}_{\sigma},
\label{VR_a}\\
{\cal L}_{\lambda}&=&\sqrt{\lambda^2+\lambda_s^2+2\lambda\lambda_s|\cos 
\phi_{12}|}
\label{VR_b}
\end{eqnarray}
with $\lambda=\{L,\sigma\}$, whereas for discorectangles
\begin{eqnarray}
V_{\rm{excl}}^{\rm{HDR}}(L_s,\sigma_s,\phi_{12})&=&(L+L_s)(\sigma
+\sigma_s)+\frac{\pi}{4}(\sigma+\sigma_s)^2 \nonumber \\&+&LL_s|\sin\phi_{12}|.
\label{VDR}
\end{eqnarray}
The reversible work required to insert the scaled particle with fixed orientation 
(which coincides with the excess chemical potential) is, in the limit of small sizes\cite{Reiss}
$(L_s,\sigma_s)\ll (L,\sigma)$:
\begin{eqnarray}
\beta \mu_{\rm{ exc}}(\phi_1)&\sim& 
\mu^{(0)}(L_s,\sigma_s,\phi_1)\nonumber \\=
&-&\ln\left[1-\rho \left<V_{\rm{excl}}\right>(L_s,\sigma_s,\phi_1)\right],
\label{magnitud}
\end{eqnarray}
where $\rho$ is the fluid density. In the opposite limit of large sizes $(L_s,\sigma_s)\gg (L,\sigma)$,
this work coincides with the thermodynamic reversible work necessary to open a cavity of
volume equal to the scaled-particle volume ($v_s$), which is equal to $Pv_s$, $P$ being the fluid
pressure. SPT interpolates between these two limits using a Taylor expansion around 
$(L_s,\sigma_s)=(0,0)$, where the second-order term is $Pv_s$ and, finally, the sizes are taken
to be those of the fluid particles, which results in
\begin{eqnarray}
\beta \mu_{\rm{exc}}(\phi_1)&=&-\ln(1-\eta)+\frac{\rho}{1-\eta}\int d\phi_2h(\phi_2)
V^{(0)}(\phi_{12})\nonumber \\&+&\beta Pv,
\label{pote}
\end{eqnarray}
where $\eta=\rho v$ is the packing fraction and $V^{(0)}=V_{\rm{excl}}(L,\sigma,\phi_{12})-2v$, with $v$ the particle volume.
The excess chemical potential of the fluid is the angular average
\begin{eqnarray}
\beta\mu_{\rm{exc}}&=&\int d\phi_1 h(\phi_1)
[\beta\mu_{\rm{exc}}(\phi_1)]=
-\ln(1-\eta)\nonumber \\&+&\frac{2\eta}{1-\eta}S_0+\beta Pv,
\label{inserta}
\end{eqnarray}
where we have defined $S_0=\langle\langle V^{(0)}\rangle\rangle/(2v)$ and
$\langle\langle \cdots\rangle\rangle$ is the double angular average. Using the definitions of
the excess chemical potential and pressure in terms of the excess free energy density, 
$\Phi_{\rm{exc}}$, a differential equation for 
$\varphi_{\rm{exc}}=\Phi_{\rm{exc}}v$ is
obtained, with solution
\begin{eqnarray}
\varphi_{\rm{exc}}
=\eta\left[-\ln(1-\eta)+\frac{\eta}{1-\eta}S_0\right]
\label{excess}
\end{eqnarray}
and, correspondingly, the pressure is given by
\begin{eqnarray}
\beta Pv=\frac{\eta}{1-\eta}+\frac{\eta^2}{(1-\eta)^2}S_0
\end{eqnarray}
which defines the thermodynamics of the fluid.

In order to proceed we have to specify a functional form for the 
orientational distribution function,
which is taken to be a truncated Fourier expansion. In view of the symmetry 
$h(\phi)=h(\pi-\phi)$, only even harmonic terms have to be included:
\begin{eqnarray}
h(\phi)=\frac{1}{\pi}\left(1+\sum_{k\ge 1}^n h_k\cos(2k\phi)\right).
\end{eqnarray}
Note that $\int_0^{\pi}d\phi h(\phi)=1$.
The cutoff index $n$ is chosen to guarantee a small enough value $|h_n|<10^{-7}$.
With this choice the double angular average $S_0$ for hard rectangles becomes
\begin{eqnarray}
S_0&=&\frac{1}{\pi}
\left(\kappa^{1/2}+\kappa^{-1/2}\right)^2
\left[1-\frac{1}{2}\sum_{k\ge 1}^n g_k h_k^2 \right],
\label{cuala}
\end{eqnarray}
where
\begin{eqnarray}
g_k&=&\left(\frac{\kappa-1}{\kappa+1}\right)^2\left(4k^2-1\right)^{-1}
, \quad  \quad k=2j+1, \nonumber\\
g_k&=&\left(4k^2-1\right)^{-1}, \quad  \quad k=2j.
\label{cuala_1}
\end{eqnarray}
We have defined $\kappa\equiv L/\sigma$ (the aspect ratio).
The equilibrium orientational structure of the fluid is obtained by minimizing the
free energy per particle, whose ideal part reads
\begin{eqnarray}
\frac{\varphi_{\rm{id}}}{\eta}=\ln \eta -1+ \int_0^{\pi}d\phi h(\phi)\ln\left[\pi h({\phi})\right]
\end{eqnarray}
whereas the excess part $\varphi_{\rm{exc}}/\eta$ is obtained from Eqn. (\ref{excess}).
A bifurcation analysis for small values of the $h_k$ coefficients leads to the following
expression for the difference between free energies $\Delta\varphi=\varphi_N-\varphi_I$
of nematic (N) and isotropic (I) phases:
\begin{eqnarray}
\frac{\Delta\varphi}{\eta}\approx \left(1-\frac{2}{3\pi}y\kappa_{-}^2\right)
\frac{h_1^2}{4}+\left(1-\frac{2}{15\pi}y\kappa_{+}^2\right)\frac{h_2^2}{4}
\end{eqnarray}
where $\kappa_{\pm}=\kappa^{1/2}\pm \kappa^{-1/2}$ and $y=\eta/(1-\eta)$. 
Note that the first term is dominant with
respect to the second. The isotropic phase gets unstable 
with respect to the uniaxial nematic phase, N$_u$, ($h_1\ne 0$) when the following condition holds:
$1-(2/3\pi)y\kappa_{-}^2=0$. Another possibility is that the isotropic phase gets unstable when 
$h_1=0$, $h_2\ne 0$ and $1-(2/15\pi)y\kappa_{+}^2=0$. This bifurcation corresponds to
a phase with D$_{4h}$ symmetry, which we call {\it tetratic}, N$_t$. The above conditions
occur for the following values of the packing fractions:
\begin{eqnarray}
\eta_{\rm{\hbox{\tiny N}}_u}=\left(1+\frac{2}{3\pi}\kappa_-^2\right)^{-1},\quad
\eta_{\rm{\hbox{\tiny N}}_t}=\left(1+\frac{2}{15\pi}\kappa_+^2\right)^{-1}
\end{eqnarray}
The equality of these two values gives a critical aspect ratio $\kappa^*=(3+\sqrt{5})/2$
such that, when $\kappa<\kappa^*$, the N$_t$ phase preempts the N$_u$ phase.

In order to calculate the complete phase diagram for uniform phases we minimized the
full free energy per particle for each $\kappa$ with respect to the $n$ coefficients $h_k$. 
A common-tangent construction was used to locate coexistence boundaries in the case
of first-order transitions. To characterize the orientational order we have used the
following order parameters:
\begin{eqnarray}
q_k=\int_0^{\pi}h(\phi)\cos (2 k\phi)d\phi, \quad k=1,2.
\label{ops}
\end{eqnarray}

\subsection{Density-functional theory for nonuniform phases}
\label{version_1}

Two different approximations are being proposed for the excess part 
of free energy density $\Phi_{\rm{exc}}({\bf r})$ which defines the interaction 
part of the 
free energy functional as $\beta {\cal F}_{\rm{exc}}=\int d{\bf r}
\Phi_{\rm{exc}}({\bf r})$. The first one will 
later be used to calculate the spinodal instabilities of the HR
fluid against nonuniform phases.

As an important ingredient of any density functional approximation 
for freely rotating anisotropic particles, we need to impose 
the Onsager form for the excess part of the free-energy 
density in the low density limit:

\begin{eqnarray}
\Phi_{\rm{exc}}^{\rm{ONS}}
({\bf r}_1)&=& -\frac{1}{2}\int d{\bf r}_2\int d\phi_1
\int d\phi_2 \rho({\bf r}_1,\phi_1)\rho({\bf r}_2,\phi_2)\nonumber \\
&\times& f({\bf r}_1,{\bf r}_2,\phi_1,\phi_2).
\end{eqnarray} 
where $f({\bf r}_1,{\bf r}_2,\phi_1,\phi_2)$ is the Mayer function 
between two particles with orientations $\phi_1$ and $\phi_2$.  
Another important feature is that the functional capture correctly 
the high-density limit. For this purpose we will use the FMT to 
include the remaining part of the density dependence and ensure,
in this way, the high density limit. This 
dependence enters through the weighted densities
$n_{\alpha}({\bf r})$\cite{Cuesta}.
Finally we impose that the scaled particle theory be recovered in the 
uniform limit.  

As a first proposal we write the following 
form for the excess part of the free energy density:
\begin{eqnarray}
\Phi_{\rm{exc}}({\bf r})&=&-n_0({\bf r})\ln[1-n_2({\bf r})]-
\frac{n_0({\bf r})n_2({\bf r})}{1-n_2({\bf r})}\nonumber \\
&+&\left[(1-n_2)^{-1}
\ast \omega^{(0)}_y\right]({\bf r})\Phi_{\rm{exc}}^{\rm{ONS}}({\bf r}),
\label{first}
\end{eqnarray}
where the $n_{\alpha}$'s are defined as $n_{\alpha}({\bf r})=\left[\rho\ast
\omega_{y}^{(\alpha)}\right]({\bf r})$, i.e. they are convolutions 
of the averaged density profile of rectangles,
$\rho({\bf r})=\int d\phi 
\rho({\bf r},\phi)$, with some weighting functions given by the expressions
\begin{eqnarray}
\omega^{(0)}_y({\bf r})&=&\frac{1}{4}\delta\left(\frac{\sigma}{2}-|x|
\right)\delta\left(\frac{L}{2}-|y|\right),\label{weight0}\\
\omega^{(2)}_y({\bf r})&=&\Theta\left(\frac{\sigma}{2}-|x|
\right)\Theta\left(\frac{L}{2}-|y|\right),\label{weight2}
\end{eqnarray}
where $\delta(x)$ and $\Theta(x)$ are the Dirac delta function and 
the Heaviside function, respectively.
An important property of the above proposal is that it recovers the
SPT theory in the uniform limit.
Note that weighting functions 
are the characteristic functions which define the support 
of the particle geometry (a rectangle parallel to the $y$ axis)  
and its total surface area  
\cite{Cuesta} (their integrals are two of the fundamental 
measures of the particles).
Note also that, in order to numerically implement the theory,
the simplest choice is to take a system of parallel HR
as a reference system. 
Another possibility is to use as a reference 
system a fluid of parallel hard squares with particles having  
the same volume as that of the rectangles. 
This choice may be more adequate to describe isotropic 
nonuniform phases such as the plastic solid (for which the lattice period
is the same in the $x$ and $y$ directions), whereas the former choice 
should be 
better for the oriented solid phases. Both reference systems will later 
be used to 
calculate the spinodal instabilities to nonuniform phases.   

A possible improvement of the theory would be to get rid of the
reference system of parallel particles and generalize the
definitions of weighting functions (\ref{weight0}) and (\ref{weight2}) 
to describe freely rotating particles:

\begin{eqnarray}
\omega_y^{(0)}({\bf r},\phi)=\frac{1}{4}\delta\left(
\frac{\sigma}{2}-|X_{\phi}|\right)\delta\left(\frac{L}{2}-|Y_{\phi}|\right), 
\\
\omega_y^{(2)}({\bf r},\phi)=\Theta\left(
\frac{\sigma}{2}-|X_{\phi}|\right)\Theta\left(\frac{L}{2}-|Y_{\phi}|\right), 
\end{eqnarray}
where $X_{\phi}=x\cos \phi+y\sin \phi$ and $Y_{\phi}=y\cos \phi-x\sin\phi$ 
are the new Cartesian coordinates in a system rotated by the local angle 
$\phi$ of the particle, measured w.r.t. the nematic director (the $y$ axis). 
Thus the weighted densities $n_{\alpha}$'s are now calculated from 
\begin{eqnarray}
n_{\alpha}({\bf r})=\int d{\bf r}'\int d\phi'\rho({\bf r}+{\bf r}',\phi')
\omega^{(\alpha)}_y({\bf r}',\phi').
\end{eqnarray}
The new expression for $\Phi_{\rm{exc}}$ is now given by 
Eqn. (\ref{first}) where the last term is substituted by
\begin{eqnarray}
&-&\frac{1}{2}\int d{\phi_1}\left[(1-n_2)^{-1}\ast\omega_y^{(0)}\right]({\bf r}_1,\phi_1)
\rho({\bf r}_1,\phi_1)\nonumber \\
&\times&\int d{\bf r}_2\int d\phi_2\rho({\bf r}_2,\phi_2)
f({\bf r}_1,{\bf r}_2,\phi_1,\phi_2),
\end{eqnarray}
This term has to be modified due to the fact that the term $(1-n_2)^{-1}$ is 
now convoluted
with the weight $\omega^{(0)}_y({\bf r},\phi)$, which explicitly depends 
on $\phi$. This approach will not be pursued any further and we leave it
for future work.

As can be seen the prefactor 
of the function $\Phi_{\rm{exc}}^{\rm{ONS}}$ in Eq. (\ref{first}) 
depends on the 
density profile defined in a rectangular region centered at ${\bf r}$ 
with dimensions $(2L)\times(2\sigma)$. This is not the usual case 
in the formulation of the FMT from first principles, but it is 
necessary because the term $\Phi_{\rm{exc}}^{\rm{ONS}}$ depends on 
the Mayer functions which have the range of the excluded surface area.  

In order to test the performance of this version of DF we will 
apply it to the study of freezing in a parallel hard-square (PHS) fluid and 
compare the results with the FMT derived in \cite{Cuesta}. This test 
is motivated by the fact that the close packing limit is easier to 
reach in this system compared with the freely rotating case and thus 
the possible divergences inherent to the particular 
functional structure can be more easily elucidated.

\begin{figure}[h]
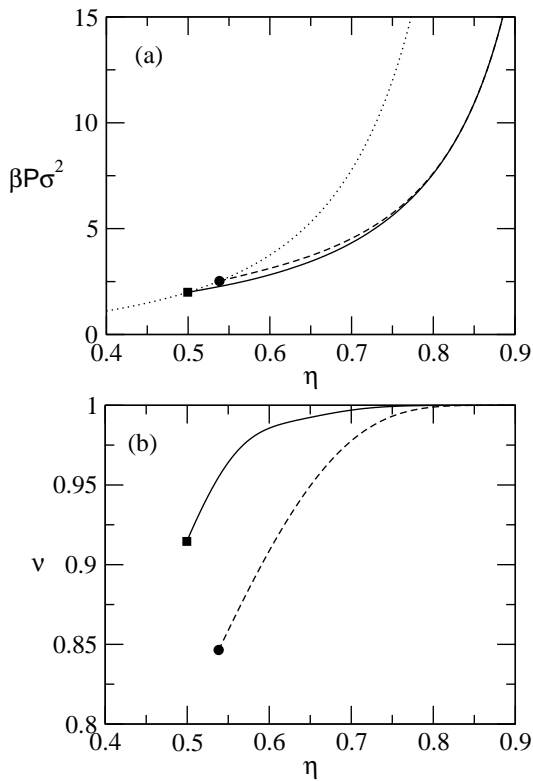

{\centering \resizebox*{7.cm}{!}{\includegraphics{fig2a.eps}} \par}
\hspace*{0.2cm}
{\centering \resizebox*{6.7cm}{!}{\includegraphics{fig2b.eps}} \par}
\caption{\label{fig0}\small (a) Equations of state for
PHS's as obtained
from FMT and the proposed version of DF. Dotted line: fluid branch;
dashed line: solid branch from FMT; and solid line: solid branch from
the proposed version. The full square and the circle indicate the
corresponding bifurcation points. (b) The occupancy probability vs
packing fraction along the solid branches. Dashed line: FMT, and
solid line: present theory.}
\end{figure}

The density profile of the solid phase was parameterized using 
Gaussian peaks centered at the sites of a square lattice and the 
fraction of vacancies was included through a normalization 
factor in the profile. The minimization of the functional was 
carried out with respect to the Gaussian width and occupancy 
probability $\nu$ (one minus the fraction of vacancies). The results 
are shown in Fig. \ref{fig0} where the fluid and solid equations of 
state (Fig. \ref{fig0}a) and the occupancy probability 
vs packing fraction (Fig. \ref{fig0}b) are plotted.
Comparison is made between results from the present theory and FMT.
The fluid branch is the same because, as already pointed
out, our theory recovers SPT. Note that the new version 
stabilizes the solid phase at a lower packing fraction 
$\eta \approx 0.50$ with a fraction of vacancies $1-\nu=0.085$, 
to be compared with the FMT result 
($\eta\approx 0.53$ and $1-\nu=0.154$). The lattice parameter, $d$, 
of the solid 
phase at the bifurcation point, as calculated from the 
mean density $\rho=\nu/d^2$, is equal to 1.353 from the present version   
and 1.254 from FMT. All of these results indicate that both functionals 
produce similar results and, what is more important, the new functional is 
regular for this kind of density profiles (note that a free 
minimization would be required to completely settle this point).

Another possibility to construct a DF for freely rotating 
hard rectangles is to impose that this functional recover the Fundamental
Measure Functional (FMF) when 
the density profile is  chosen to have the form 
$\rho({\bf r},\phi)=
\rho_{y}({\bf r})\delta(\phi)$, i.e. the system of parallel hard rectangles. 
Note that, in this limit, we should recover the equation of state of the solid
corresponding to a system of parallel hard squares, shown in Fig. \ref{fig0}a.
A free-energy density that conforms to this criterion is proposed in the
Appendix.

\subsection{Bifurcation analysis of nonuniform phases}
\label{bif}
At fixed chemical potential $\mu_0$ corresponding to a nematic phase
characterized by the one particle distribution function 
$\rho_0h(\phi)$, the minimization of the grand potential leads to 

\begin{eqnarray}
\rho({\bf r},\phi)=\rho_0h(\phi)\exp{\left[-\beta\left(
\frac{\delta {\cal F}_{\rm{exc}}}
{\delta \rho({\bf r},\phi)}-\mu_0\right)\right]}.
\label{bifurca}
\end{eqnarray}

Assuming a density profile of the form $\rho({\bf r},\phi)=\rho_0
h(\phi)+\xi({\bf r},\phi)$ for the incipient nonuniform phase 
which bifurcates from the nematic phase, a Taylor expansion of 
Eq. (\ref{bifurca}) up to first order in $\xi$ gives 

\begin{eqnarray}
&&\rho({\bf r},\phi)=\rho_0h(\phi)\nonumber \\&\times&
\left[1+\int d{\bf r}'\int d\phi'
C({\bf r}-{\bf r}',\phi,\phi')\xi({\bf r}',\phi')\right], 
\label{ecua0}
\end{eqnarray}
where $C({\bf r}-{\bf r}',\phi,\phi')$ is the direct correlation function
of the nematic fluid calculated as minus the second functional derivative of 
$\beta{\cal F}_{\rm{exc}}$ with respect to the density profile evaluated
at $\xi=0$. 
Eqn. (\ref{ecua0}) can be rewritten in Fourier 
space as 

\begin{eqnarray} 
\hat{\xi}({\bf q},\phi)-\rho_0h(\phi)\int d\phi'\hat{C}({\bf q},\phi,\phi')
\hat{\xi}({\bf q},
\phi')=0.
\label{Fourier} 
\end{eqnarray}
where the hats over the functions $\xi$ and $C$ indicate Fourier transforms.

Inserting the truncated Fourier expansion of the
function $\hat{\xi}({\bf q},\phi)$,
\begin{eqnarray} 
\hat{\xi}({\bf q},\phi)=\sum_{k=0}^n\xi_k({\bf q})\cos(2k\phi)
\label{Fourier1} 
\end{eqnarray}
in Eqn. (\ref{Fourier}), multiplying 
the latter by $\cos(2j\phi)$ and integrating in $\phi$ from 0 to 
$2\pi$ we obtain the following algebraic equation: 

\begin{eqnarray}
\left(I-\rho_0 T\right){\bf u}=0, 
\label{algebraica}
\end{eqnarray}
where $I$ is the identity matrix, $T$ is a $(n+1)\times(n+1)$ matrix
with elements
\begin{eqnarray}
T_{jk}&=&\frac{1}{\pi(1+\delta_{j0})}\int d\phi h(\phi)\nonumber \\
&\times&\int d\phi'\cos(2j\phi)
\cos(2k\phi')\hat{C}({\bf q},\phi,\phi'),
\label{elementos}
\end{eqnarray}
and ${\bf u}=[\xi_0({\bf q}),\cdots,\xi_n({\bf q})]^T$ is 
a $n+1$ vector.

The lower value of $\rho_0$ and the vector ${\bf q}$ 
for which there exists a non trivial solution to Eqn. (\ref{algebraica}) can 
be calculated from 

\begin{eqnarray}
H(\rho_0,{\bf q})\equiv\hbox{det}\left(I-\rho_0T\right)&=&0, \label{bifurca_1}\\ 
\boldsymbol{\nabla}H(\rho_0,{\bf q})&=&0 \label{bifurca_2}
\end{eqnarray}
Both equations will be used later to calculate the spinodal instabilities  
from the isotropic or nematic phases to possible nonuniform phases.

\section{Results} 

\subsection{Uniform phases}
\begin{figure}[h]
{\centering \resizebox*{7cm}{!}{\includegraphics{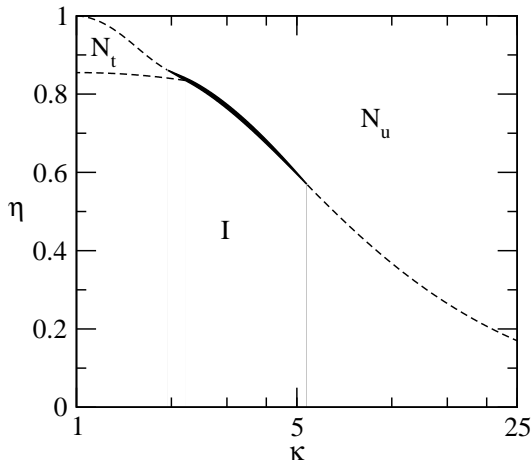}} \par}
\caption{\label{fig1}\small Phase diagram for hard rectangles in the $\eta-\kappa$ plane.
The $\kappa$ axis is represented using a logarithmic scale.
Dashed lines indicate second-order transitions, whereas full lines are binodals. The shaded
region is the two-phase region where phase separation occurs. Labels indicate the three
phases involved: isotropic (I), uniaxial nematic (N$_u$) and tetratic nematic (N$_t$).}
\end{figure}

\begin{figure}[h]
{\centering \resizebox*{7cm}{!}{\includegraphics{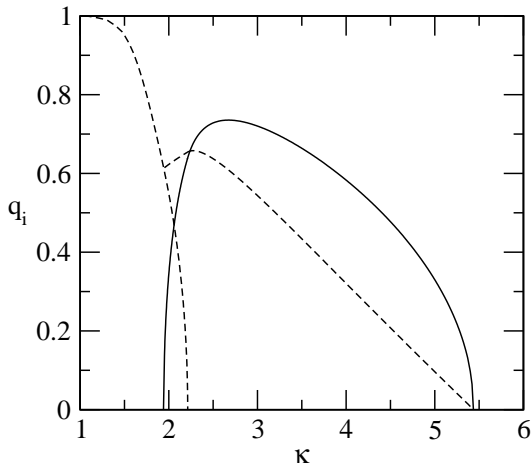}} \par}
\caption{\label{fig2}\small Order parameters $q_1$ and $q_2$ along the transition lines
to orientationally ordered phases. The full line corresponds to the $q_1$ parameter, whereas
the dashed lines correspond to the $q_2$ parameter.}
\end{figure}

\begin{figure}[h]
{\centering \resizebox*{7cm}{!}{\includegraphics{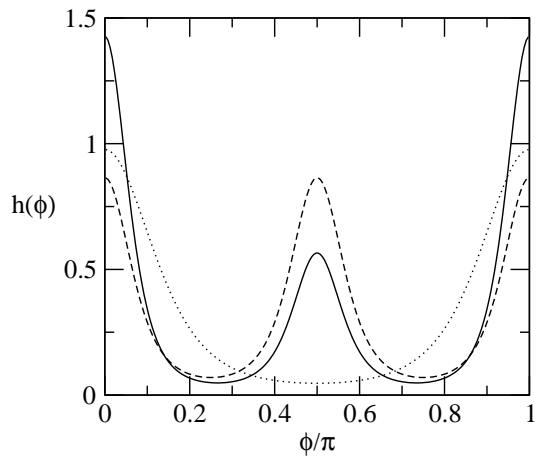}} \par}
\caption{\label{fig3}\small Angular distribution
functions of hard rectangles
of the N$_u$ (solid line) and N$_t$ (dashed line) phases at coexistence
and for $\kappa=2$. The dotted line represents
the distribution function of hard discorectangles
for the same $\kappa$ and for $\eta=0.91$.}
\end{figure}

In Fig. \ref{fig1} we show the phase diagram for hard rectangles with $1\le\kappa\le 25$
as obtained by applying the SPT formalism. For 
$2.21<\kappa<5.44$ the I$-$N$_u$ transition is of first order. For $\kappa<2.21$
a tetratic nematic phase N$_t$ begins to be stable; its stability region is bounded
below by a second-order I$-$N$_t$ transition. The region is bounded above 
by a N$_t-$N$_u$ first-order transition for $1.94<\kappa<2.21$ and by a second-order
transition for $1<\kappa<1.94$. This means that there exist two tricritical points: one 
at $\kappa=5.44$, already predicted by Schlacken et al.\cite{Schlacken},
and a second one at $\kappa=1.94$, which can only be predicted by a proper bifurcation analysis
from the N$_t$ to the N$_u$ phases.
Fig. \ref{fig2} shows the two order parameters [Eqns. (\ref{ops})] 
along the transition lines to orientationally ordered phases, 
for the systems where these parameters are 
different from zero.  
Orientational distribution functions for the N$_u$ and N$_t$ phases at coexistence
for $\kappa=2$ are shown in Fig. \ref{fig3}. Note that the N$_t$ phase presents
two peaks of the same height (consistent with this phase being invariant with respect
to the D$_{4h}$ symmetry group). By contrast, the N$_u$ phase also presents a secondary
peak (indicating a certain degree of biaxiality in this phase), but of a smaller height. If $\kappa$ is
increased the height of this secondary peak reduces compared to the main peak, eventually
disappearing.

We now show the results for hard discorectangles. We have implemented the same calculational
scheme and obtained the phase diagram. Only two phases are involved: the isotropic phase I
and the uniaxial nematic N phase. The phase boundary $\eta-\kappa$ between the two turns out to be of
second order always\cite{comment}, and is given by the equation
\begin{eqnarray}
\eta=\left(1+\frac{2}{3}\frac{\displaystyle(\kappa-1)/\pi}{\displaystyle
1+\frac{\pi}{4(\kappa-1)}}
\right)^{-1},
\end{eqnarray}
where $\kappa=(L+\sigma)/\sigma$ and now $L$ is the length of the rectangular part of the
particle and $\sigma$ is its width. 

A key difference between hard rectangles and hard discorectangles is that in the former
case the excluded volume expression contains a cosine term which is absent in the latter
case. This can be seen from Eqns. (\ref{VR_a}), (\ref{VR_b}) and 
(\ref{VDR}); substituting $L_s=L$ and
$\sigma_s=\sigma$ we obtain
\begin{eqnarray}
V_{\rm{excl}}^{\rm{HR}}(L,\sigma,\phi_{12})&=&
(L^2+\sigma^2)|\sin \phi_{12}|\nonumber \\&+&
2L\sigma\left(1+|\cos \phi_{12}|\right)
\end{eqnarray}
whereas for discorectangles
\begin{eqnarray}
V_{\rm{excl}}^{\rm{HDR}}(L,\sigma,\phi_{12})=4L\sigma
+\pi\sigma^2+L^2|\sin\phi_{12}|.
\end{eqnarray}

This basic difference translates into the $g_k$ prefactors in Eqn. (\ref{cuala})
being different according to the parity of the index for hard rectangles 
[see Eqn. (\ref{cuala_1})],
whereas for hard discorectangles these coefficients have the same form. This feature again
shows up in the orientational distribution function which, as pointed out above, 
exhibits a secondary maximum at $\phi=\pi/2$ in the case of rectangles,
a peak which is absent in the case of discorectangles,
see Fig. \ref{fig3}. In this figure we also show the angular distribution
function corresponding to the N phase of hard discorectangles
for the same $\kappa$ (and for $\eta=0.91$). 

\begin{figure}[h]
{\centering \resizebox*{7cm}{!}{\includegraphics{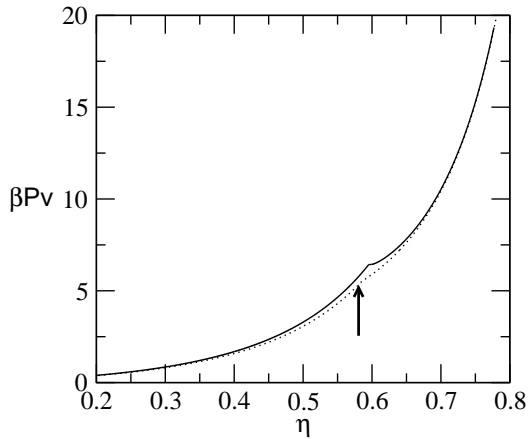}} \par}
\caption{\label{fig4}\small Equations of state of hard rectangles (
solid line) and
discorectangles (dotted line) for $\kappa=5$. The arrow indicates the
location of the second order phase transition.}
\end{figure}

In order to compare the equations of state of hard rectangles and 
discorectangles for the same aspect ratio $\kappa=5$,
the reduced pressure vs packing fraction is shown
in Fig. \ref{fig4}. It can be seen that the two equations of state
show a similar behavior, except in the neighborhood of the corresponding
phase transitions, which are of first and second order for rectangles
and discorectangles, respectively.

\subsection{Nonuniform phases}

\begin{figure}[h]
{\centering \resizebox*{7.cm}{!}{\includegraphics{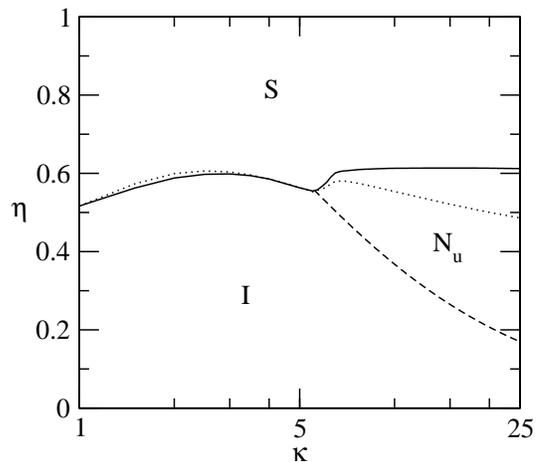}} \par}
\caption{\label{fig5}\small Phase diagram of hard rectangles including
the spinodal lines of the I-S and N$_{\rm{u}}$-S transitions. The solid
and dotted lines represent the spinodal curves calculated from the functional
obtained from Eqn. (\ref{first}), using a reference system
of parallel rectangles and parallel squares, respectively.
For the latter case the volume of the squares were taken to be the same as that 
of rectangles.}
\end{figure}

To elucidate the absolute stability of the tetratic nematic phase  
(see Fig. \ref{fig1}), 
we have carried out a bifurcation 
analysis following the lines described in section \ref{bif}, i.e. solving 
Eqns. (\ref{bifurca_1}) and (\ref{bifurca_2}) in order to find the 
values of the packing fraction and wave vector for which 
the uniform phases destabilize to 
nonuniform phases. The results are shown in Fig. \ref{fig5} where we
plot all spinodal lines of the I-N$_{\rm{u}}$, the 
I-S (S being a spatially ordered phase, whose structure will require further analysis) 
and the N$_{\rm{u}}$-S transitions. 
Note that Fig. \ref{fig5} is in fact the same as Fig. \ref{fig1}, but spinodal lines 
from the above bifurcation analysis have been superimposed and, at the same
time, transition lines in Fig. \ref{fig1} which turn out to be preempted by 
nonuniform phases have been dropped. 

As can be seen in Fig. \ref{fig5}, for $\kappa\lsim 5.5$ the I phase  
destabilizes directly to the S phase, so the tetratic nematic phase is 
not stable. Nevertheless, a tetratic order is possible in spatially ordered 
phases. Taking into account that the I-S transition is probably of
first order a complete functional minimization with respect to 
two dimensional density profiles is required to settle this question. 
In Fig. \ref{fig5} the nonuniform phase spinodals are calculated using two 
reference systems: parallel rectangles and parallel squares.
As was discussed in section \ref{version_1}, the last one 
should better describe the I-S 
transition, whereas the former should be more appropriate for the N$_{\rm{u}}$-S transition.
In other words, the dotted line should be more reliable for 
$\kappa <5.5$, whereas the solid line is expected to be more accurate for $\kappa>5.5$
(note that both reference systems give similar results in the first region and naturally
these results coincide exactly at $\kappa=1$).

If we use the density functional
for freely rotating particles without a reference system of parallel particles (see section
\ref{version_1}), the fluid-solid transition  
is expected to shift to higher densities. The reason for this behavior is that a reference system composed 
of parallel particles enhances a crystalline local order which is absent in the isotropic 
phase. Nevertheless, we expect that, within the present version of the theory, the tetratic phase 
still remains unstable w.r.t. 
the solid phase. Calculations using this functional is a task in progress and will allow 
us to finally settle this question.

\section{Conclusions}

The main conclusion of the present work is that different particle
geometries may account for the different nature of the orientational
phase transitions involved (continuous versus first order 
with a tricriticcal point). But, even
more important, particles possessing uniaxial symmetry (hard rectangles)
may lead to a phase with a dramatically different symmetry: the tetratic
phase. This finding is based upon minimization of a density-functional
theory constructed using ideas from SPT. However, the tetratic phase
has been found to be only metastable with respect to a spatially ordered phase.
This result was obtained using a bifurcation analysis of a version
of density-functional theory proposed in the present paper. The theory
combines the Onsager functional with the FMF for hard rectangles. A
word of caution is in order here: the proposed functional does not 
reproduce the exact third virial coefficient, which is expected to
be of vital importance at high density, and these neglected correlations
should play a role of paramount importance in the stabilization of
a phase with tetratic symmetry. It might well be that a full theory,
incorporating correlations between three particles, could stabilize
the tetratic phase even further and predict it to be stable at densities
below that at which crystallization occurs. Nevertheless, it is interesting
to note that the present theory already confirms that an oriented fluid
of hard rectangles presents some degree of biaxial order for small
aspect ratios. The theory could also be used to predict the stability
of a number of two-dimensional phases that occur in Langmuir 
monolayers\cite{Kaganer}.

Previous work on polydisperse mixtures of hard spherocylinders 
\cite{Polson, Cinacchi_2}, or  parallelepipeds\cite{Martinez-Raton_3}
in three dimensions, have shown that polydispersity enhances the stability of the uniform phases 
w.r.t. nonuniform phases. While columnar order admits a higher degree 
of polydispersity, the smectic or solid phases normally have a terminal polydispersity 
beyond which these phases are no longer stable\cite{Polson}. This fact can be easily understood if we 
take into account that: i) any periodic packing is difficult to attain in a mixture
of  particles with different characteristic lengths and ii) mixing increases 
the ideal mixing entropy of uniform phases and, as a consequence, decreases the Helmholtz free energy. 
Thus, some degree of polydispersity 
(either unimodal or bimodal) will certainly increase the density of the transition from the fluid
to a spatially ordered phase in a 
polydisperse mixture of hard rectangles. Is this effect enough to stabilize the 
tetratic phase? This is an open question that we will try to elucidate in a future work.   

\begin{acknowledgments}
Y.M.-R. was supported by a Ram\'on y Cajal research contract. This work is part of research
projects Nrs. BFM2003-0180, BFM2001-0224-C02-01, BFM2001-0224-C02-02 and BFM2001-1679-C03-02 (DGI) of 
the Ministerio de Educaci\'on y Ciencia (Spain).

\end{acknowledgments}

\section*{Appendix}

In this Appendix we present an alternative version of DF which 
recovers the FMT for parallel HR and SPT in the uniform limit.
Our proposal for the excess free-energy density is
\begin{eqnarray}
\Phi_{\rm{exc}}({\bf r})=-n_0({\bf r})\ln[1-n_2({\bf r})] -
\frac{n_{1x}({\bf r})n_{1y}({\bf r})}{1-n_2({\bf r})}\nonumber \\
-2\sum_{\mu=x,y}\left[\frac{\partial\Phi_{\rm{exc}}^{\rm{ref}}}
{\partial n_{1\mu}}
\ast\omega^{(1\mu)}_y\right]({\bf r})
\frac{
\Phi_{\rm{exc}}^{\rm{ONS}}
({\bf r})}{\int d{\bf r}'\rho({\bf r}')f_{\rm{ref}}
({\bf r}-{\bf r}')}, 
\label{cosa}
\end{eqnarray}
where $f_{\rm{ref}}({\bf r})=-\Theta(\sigma-|x|)\Theta(L-|y|)$ is the Mayer function of two parallel 
hard rectangles,
$\Phi^{\rm{ref}}_{\rm{exc}}$ is the excess part 
of the Helmholtz free-energy density obtained from the FMT for 
parallel hard rectangles \cite{Cuesta}, i.e. 
\begin{eqnarray}
\Phi_{\rm{exc}}^{\rm{ref}}=-n_0\ln(1-n_2)+\frac{n_{1x}n_{1y}}{1-n_2},
\end{eqnarray}
while the new weights are 
\begin{eqnarray}
\omega_{y}^{(1x)}&=&\frac{1}{2}\delta\left(\frac{\sigma}{2}-
|x|
\right)\Theta\left(\frac{L}{2}-|y|\right),\\
\omega_{y}^{(1y)}&=&\frac{1}{2}\Theta\left(\frac{\sigma}{2}-
|x|
\right)\delta\left(\frac{L}{2}-|y|\right),
\end{eqnarray}
Note that the uniform limit of $\Phi_{\rm{exc}}({\bf r})v$ 
[see Eqn. (\ref{cosa})] coincides
with the expression (\ref{excess}) obtained from the SPT. 

Now we will show that the free-energy functional obtained from
Eqn. (\ref{cosa}) reduces to the FMT for the case of parallel
rectangles.
In this limit it is easily shown that
\begin{eqnarray}
\frac{
\Phi_{\rm{exc}}^{\rm{ONS}}({\bf r})
}
{\int d{\bf r}'\rho({\bf r}')f_{\rm{ref}}({\bf r}-{\bf r}^{\prime})}
\longrightarrow -\frac{1}{2}\rho({\bf r})
\end{eqnarray}
The above limit, together with the fact that
\begin{eqnarray}
\int d{\bf r}\left[\frac{n_{1\mu}}{1-n_2}\ast\omega^{(1\nu)}_y\right]
({\bf r})\rho({\bf r})=\int d{\bf r}\frac{n_{1\mu}({\bf r})n_{1\nu}
({\bf r})}{1-n_2({\bf r})},
\end{eqnarray}
(where $\mu\ne\nu$) allow us to conclude that 
\begin{eqnarray}
\int d{\bf r}\Phi_{\rm{exc}}({\bf r})\longrightarrow \int d{\bf r}
\Phi_{\rm{exc}}^{\rm ref}({\bf r}).
\end{eqnarray}


\begin{references}
\bibitem{Freiser} M. J. Freiser, Phys. Rev. Lett. {\bf 24}, 1041 (1970).
\bibitem{Madsen} L. A. Madsen, T. J. Dingemans, M. Nakata, and E. T. Samulski, 
Phys. Rev. Lett. {\bf 92}, 145505 (2004).
\bibitem{Acharya} B. R. Acharya, A. Primak, and S. Kumar,
Phys. Rev. Lett. {\bf 92}, 145506 (2004).
\bibitem{Schlacken} H. Schlacken, H.-J. Mogel, and P. Schiller, Mol. Phys. 
{\bf 93}, 777 (1998).
\bibitem{Cuesta0} J. A. Cuesta and D. Frenkel, Phys. Rev. A {\bf 42}, 2126 (1990).
\bibitem{Bates} M. A. Bates and D. Frenkel, J. Chem. Phys. {\bf 112}, 
10034 (2000).
\bibitem{Lagomarsino} M. C. Lagomarsino, M. Dogterom, and M. Dijkstra, 
J. Chem. Phys. {\bf 119}, 3535 (2003).
\bibitem{Frenkel} D. Frenkel and R. Eppenga, Phys. Rev. A {\bf 31}, 
1776 (1985).
\bibitem{Somoza}A. M. Somoza and P.Tarazona, Phys. Rev. Lett. {\bf 61},
2566 (1988); Phys. Rev. A {\bf 41}, 965 (1990).
\bibitem{Poniewierski} A. Poniewierski and R. Holyst, Phys. Rev. Lett. 
{\bf 61}, 2461 (1988).
\bibitem{Tjipto} B. Tjipto-Margo and G. Evans, Mol. Phys. {\bf 74}, 85 
(1991).
\bibitem{Roij}R. van Roij, P. Bolhuis, M. Bulder, and D. Frenkel, 
Phys. Rev. E {\bf 52}, R1277 (1995).
\bibitem{Graf} H. Graf and H. Lowen, J. Phys.: Condens. Matter {\bf 11}, 
1435 (1999).
\bibitem{Velasco} E. Velasco, L.Mederos, and D. E. Sullivan, 
Phys. Rev. E {\bf 62}, 3708 (2000).
\bibitem{Cotter}M. A. Cotter and D. Wacker, Phys. Rev. A {\bf 18}, 2669 
(1978).
\bibitem{Barboy} B. Barboy and M. Gelbart, 
J. Chem. Phys. {\bf 71}, 3053 (1979).
\bibitem{Cuesta} J.A. Cuesta and Y. Mart\'{\i}nez-Rat\'on, Phys. Rev. Lett. 
{\bf 78}, 3681 (1997). 
\bibitem{Martinez-Raton} Y. Mart\'{\i}nez-Rat\'on, Phys. Rev. E 
{\bf 69}, 061712 (2004).
\bibitem{Cinacchi} G. Cinacchi and F. Schmid, J. Phys.: Condens. Matter {\bf 14}, 12223 (2002).
\bibitem{Reiss}H. Reiss, H. L. Frisch, and J. L. Lebowitz, J. Chem. Phys. 
{\bf 31}, 369 (1959).
\bibitem{comment} Note that the general Landau expansion of the free energy
in two dimensions contains only powers of Tr $Q^2$, where $Q$ is the tensor
order parameter. Depending on the sign of the coefficient associated with the
term $\left(\hbox{Tr} Q^2\right)^2$, which in turn is determined by the
particle geometry, the transition can be of first or
second order. By contrast, in three dimensions the transition must be
of first order due to the presence of a term Tr $Q^3$. 

\bibitem{Kaganer} V. M. Kaganer, H. M\"ohwald, and P. Dutta, 
Rev. Mod. Phys. {\bf 71}, 779 (1999).
\bibitem{Polson} J.\ M.\ Polson and D.\ Frenkel, Phys.\ Rev.\ E {\bf 56},
R6260 (1997).
\bibitem{Cinacchi_2} G. Cinacchi, L. Mederos, and E. Velasco, J. Chem. Phys. {\bf 121}, 3854 
(2004).
\bibitem{Martinez-Raton_3} Y. Mart\'{\i}nez-Rat\'on and J. A. Cuesta, J. Chem. Phys. {\bf 118}, 
10164 (2003).
\end{references}
\end{document}